\documentclass[useAMS,psfig]{mn2e}
\input{psfig}

\usepackage[dvips]{graphicx}
\title{Radio-quiet objects in the 2QZ survey}
\author[M.Wals et al.]
	{M.Wals$^{1}$, B.J.Boyle$^{1}$, S.M. Croom$^{2}$ L.Miller$^{3}$, R.Smith$^{4}$, T.Shanks$^{5}$, P.Outram$^{5}$\\
$^{1}$ Australia Telescope National Facility, PO Box 76, Epping NSW, 1710 \\
$^{2}$ Anglo-Australian Observatory, PO Box 296, Epping, NSW 1710, Australia \\
${^3}$ Department of Physics, Oxford University, 1 Keble Road, Oxford, OX1 3RH \\
${^4}$ Liverpool John Moores University, Edgerton Wharf, Liverpool \\
${^5}$ Department of Physics, University of Durham, South Road, Durham, DH1 3LE\\
}

\begin{document}
\maketitle
%\date{Accepted 2003 December 15. Received 2003 December 14; in original form 2004 April 15}

\pagerange{\pageref{firstpage}--\pageref{lastpage}} \pubyear{2004}

\label{firstpage}

\begin{abstract}
Co-addition of blank-field FIRST data at the location of over 8000 QSOs in the 2QZ survey has yielded statistical detections of radio quiet QSOs with median flux levels of $20-40\mu$Jy. We show that the total radio flux of radio-quiet QSOs in the 2QZ is consistent with a smooth extrapolation of the 2QZ radio-loud QSO number-flux distribution based on the slope of the relation flattening
near the FIRST flux limit.  However, we are unable to distinguish a smooth extrapolation of the luminosity function to faint levels from a bimodal luminosity function with a break close to or below the FIRST radio detection limit. We also demonstrate that the redshift dependence of the median radio-to-optical spectral index $\alpha_{\rm RO}$ for these radio quiet QSOs is consistent with that obtained for individual radio-loud 2QZ QSOs detected by FIRST.  
\end{abstract}

%%%%%%%%%%%%%%

\section{Background}

The debate over whether the observed properties of radio-loud and radio-quiet QSOs are consistent with a single population of objects, albeit with a broad range in radio properties, or two distinct populations is more than thirty years old (see e.g. Schmidt 1970).  Evidence for bi-modality in the radio properties of QSOs claimed by a number of authors (see e.g. Kellerman et al.\ 1989, Miller, Peacock \& Mead 1990) has been challenged more recently using data obtained from the deeper, more extensive radio and optically-selected catalogues (White et al.\ 2000, Hewett, Foltz \& Chaffee 2001) that are increasingly becoming available.  However, recent claims for bi-modality have also been made based on QSO samples drawn from the extensive Sloan Digital Sky Survey (Ivesic et al. 2002, although see Ivesic et al.\ 2004 for a reassessment of this result) In contrast, a statistical analysis of the 2dF QSO redshift survey (2QZ, see Croom et al.\ 2004) and LBQS optically-selected survey and FIRST radio surveys has yielded no evidence for any gap in the distribution between radio-loud and radio-quiet populations (Cirasuolo et al. 2003).  

Due to the instrincally low radio flux of radio-quiet QSOs $\ll 100\mu$Jy, most previous analyses have been based on statistical arguments applied to upper flux limits inferred from non-dectections, as opposed to detection, of radio quiet QSOs. In this paper, we adopt a different approach by directly trying to obtain a detection for an `ensemble' average of radio-quiet QSOs in the 2dF QSO survey by stacking radio data from the FIRST survey.  The combination of the 2QZ sample size ($\sim 10000$ QSOs in the area of overlap with FIRST) and the moderate depth of the FIRST survey (0.15mJy rms) suggest it should be possible to achieve reliable detections ($>5\sigma$) for a number of 1000-2000 QSO ensembles at a level of a few tens of $\mu$Jy.  Indeed, this method has already been successfully pioneered on the Sloan QSO sample by Glickman et al.\ (2004). 

We describe the data and method used in Section 2 and present our analysis of the radio-quiet QSO flux distribution derived directly from the coadded images in Section 3. `Standard' cosmological parameters: $\Omega_{\rm m}=0.3$, $\Omega_\Lambda=0.7$ and $H_0=70\,$km$\,$s$^{-1}$Mpc$^{-1}$ are used throughout this paper.

\section{method}

We used the 2QZ to obtain accurate (0.2-0.3\,arcsec) positions for 8741 optically-selected QSOs within the 75-degree $\times$ 5-degree wide 2QZ equatorial strip ($9^{\rm h}40^{\rm m}<{\rm RA}<14^{\rm h}40$) that overlaps with the FIRST survey (White et al.\ 2001).  To ensure maximum data integrity, we used only QSOs with ID and redshift quality 1 in the 2QZ, and restricted our analysis to QSOs in the redshift range $z<2.3$ where the survey completeness is over 50 per cent (Croom et al.\ 2004). We did not exclude the lowest redshift QSOs ($z<0.25$) from the analysis where the completeness is also less than 50 per cent.  However these objects only comprise 3 per cent of the QSOs in the lowest redshift bin under study here, and we have verified that their inclusion/exclusion makes no difference to the results obtained below.   

Out of this sample of QSOs, 228 are listed in the 2QZ public release catalogue as having NVSS counterparts. A further 134 were also found to have FIRST radio counterparts; defined as a source listed in the FIRST catalogue (i.e. $>5\sigma$ detection) within 5\,arcsec of the optical position. The FIRST radio fluxes for all these 362 QSOs, hereinafter referred to as radio-loud, lay in the range 0.65mJy$<f(1.4{\rm GHz})<3092\,$mJy.  All 2QZ objects identified as a radio-loud quasar were consistent with a point source radio detection.

For the remaining 8379 QSOs we extracted 30 arcsec $\times$ 30 arcsec images centred on the optical position from the FIRST catalgoue.  Each `cut out' was visually inspected, and 70 were rejected from further analysis based on either poor data (sidelobes etc.) or the presence of a bright source in the edge of the field.  The remaining 8309 QSOs, hereinafter radio quiet QSOs, were then grouped into five redshift bins.  Each bin was designed to have an equal number of QSOs, so that the rms noise in each stacked image would be similar. The redshift bins and the numbers of QSOs are listed in Table~\ref{tab:prop}.  In practice, 
there is a small variation in the numbers of QSOs between each bin, due to an iteration in the removal of maps with poor data.  

\begin{table}
\centering
\caption{Radio properties of composite QSOs}
\label{tab:prop}
\begin{tabular}{@{}lcccc@{}}
\hline
z range&$z_{\rm med}$&$N_{\rm QSO}$&$f(\rm 1.4GHz)_{\rm med}$&rms\\
&&&($\mu$Jy)&($\mu$Jy)\\
\hline
$0.12\leq z<0.85$ & 0.65 & 1666 & 41.8 & 8\\
$0.85\leq z<1.23$ & 1.01 & 1665 & 29.8 & 7\\
$1.23\leq z<1.55$ & 1.39 & 1666 & 23.2 & 6\\
$1.55\leq z<1.89$ & 1.72 & 1666 & 28.6 & 6\\
$1.89\leq z<2.30$ & 2.05 & 1646 & 31.3 & 5\\
\hline
\end{tabular}
\end{table}

\begin{figure}
\centering
\caption{Images from co-added redshift bins.  Increasing redshift bins run from left to right and top to bottom.  The image in the lower right part of the figure is the sum of all redshift bins.}
\label{fig:bins}
\includegraphics[scale=0.12]{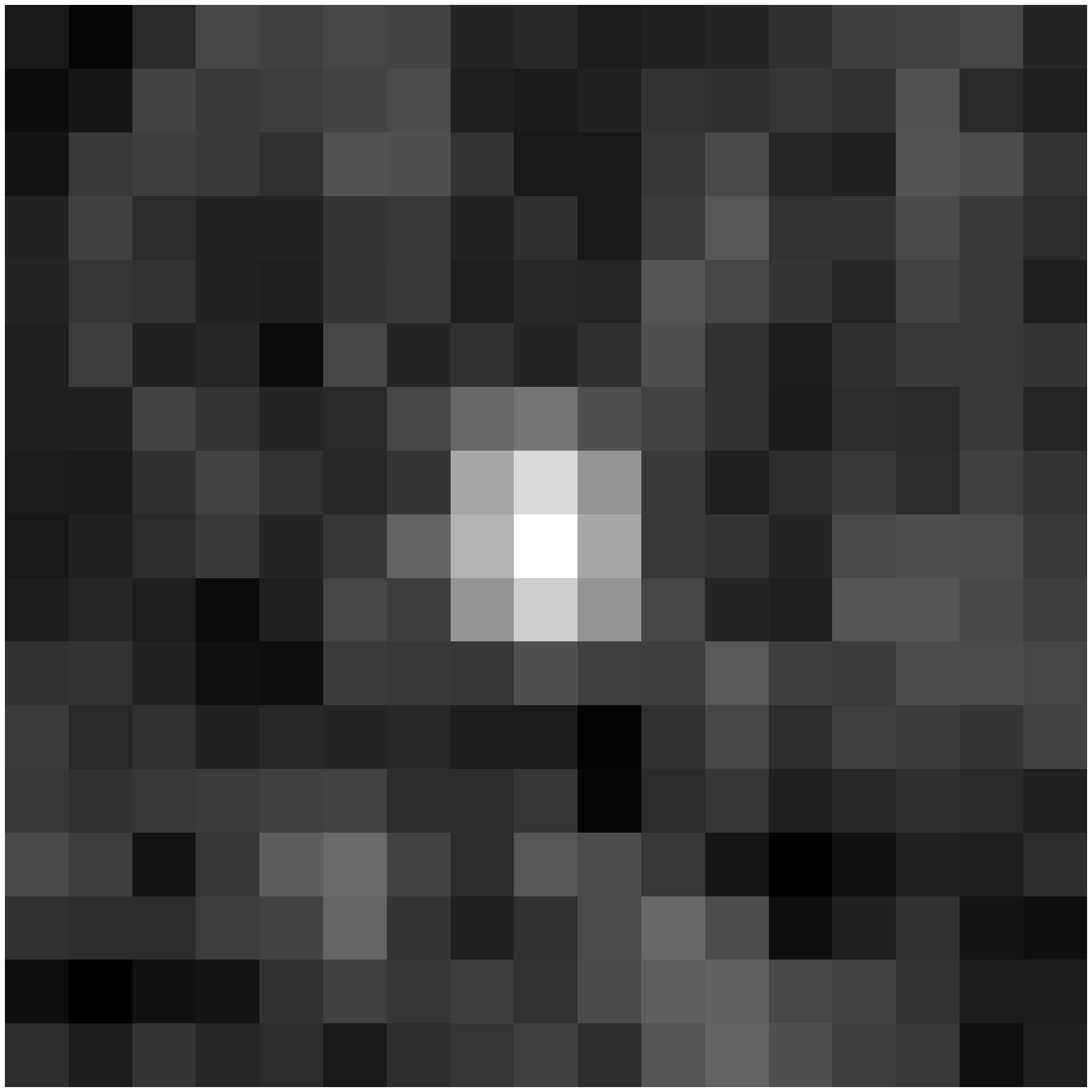}
\includegraphics[scale=0.12]{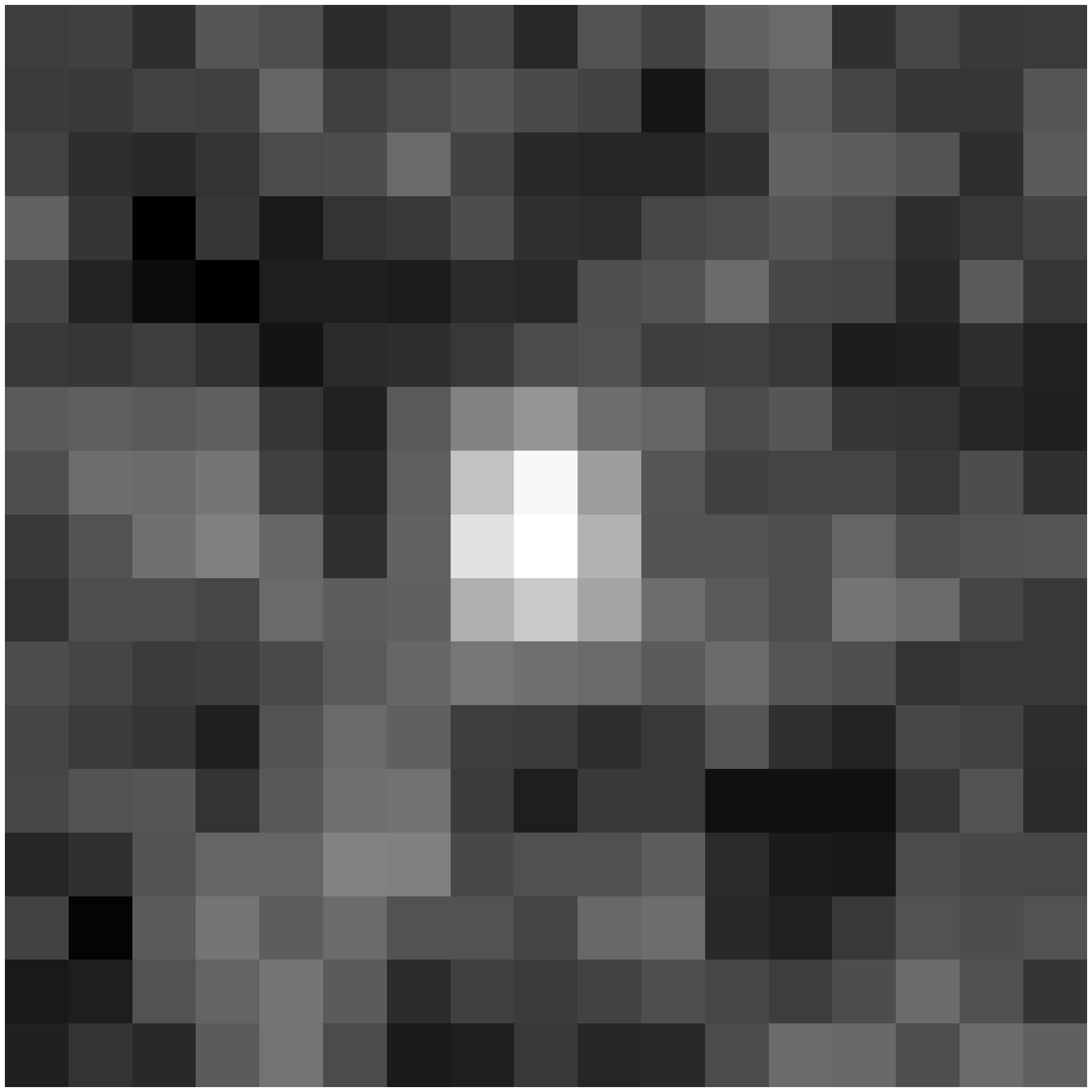}
\includegraphics[scale=0.12]{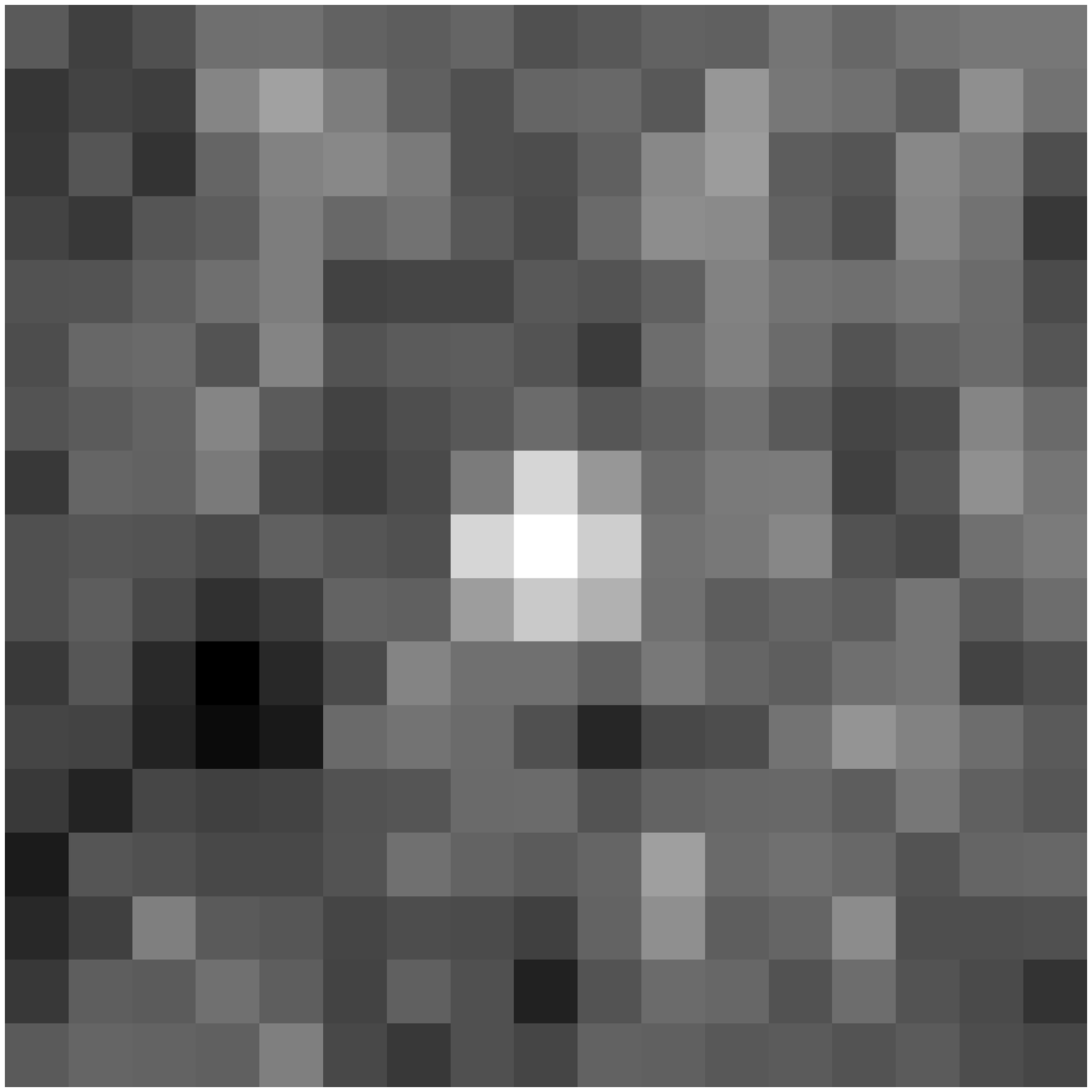}
\includegraphics[scale=0.12]{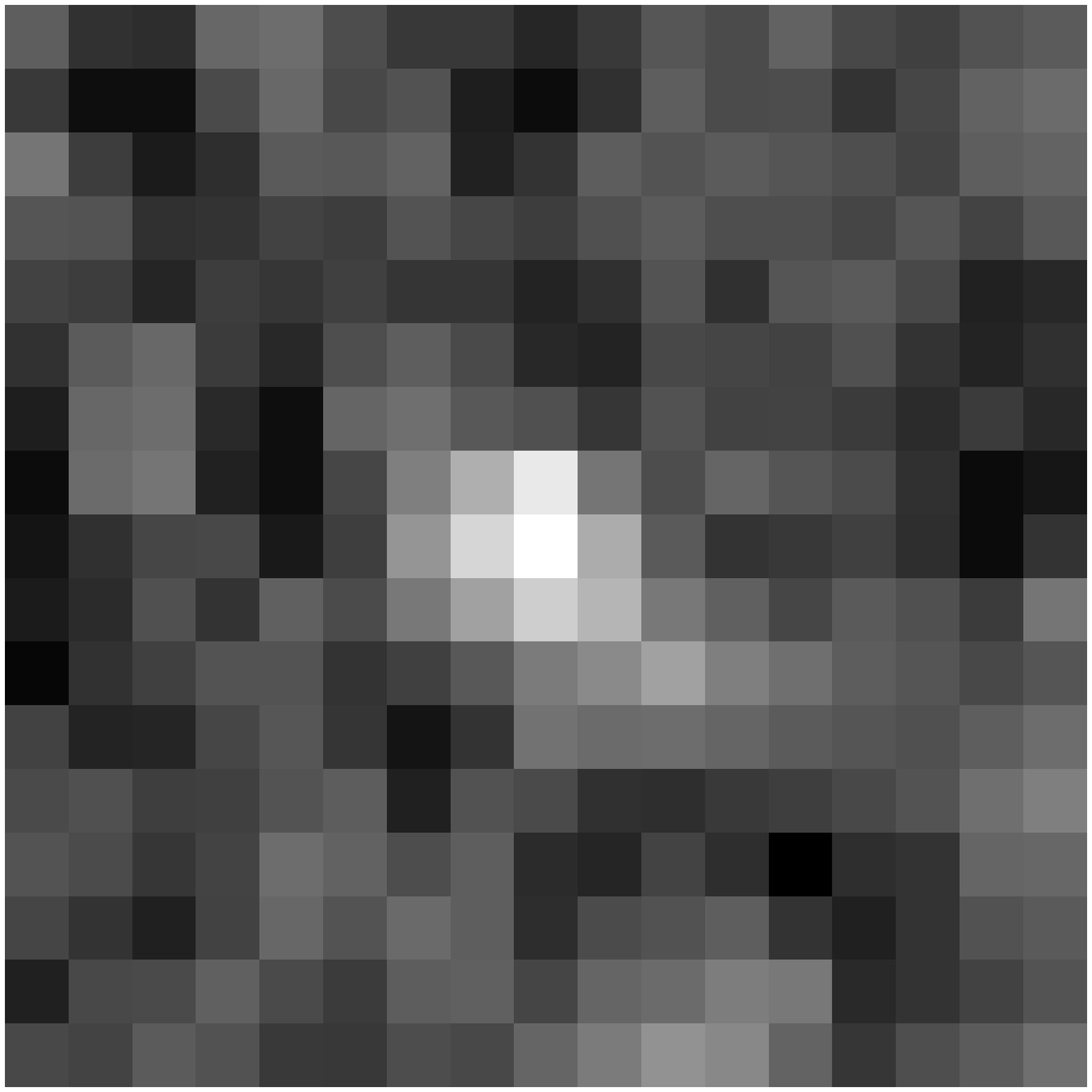}
\includegraphics[scale=0.12]{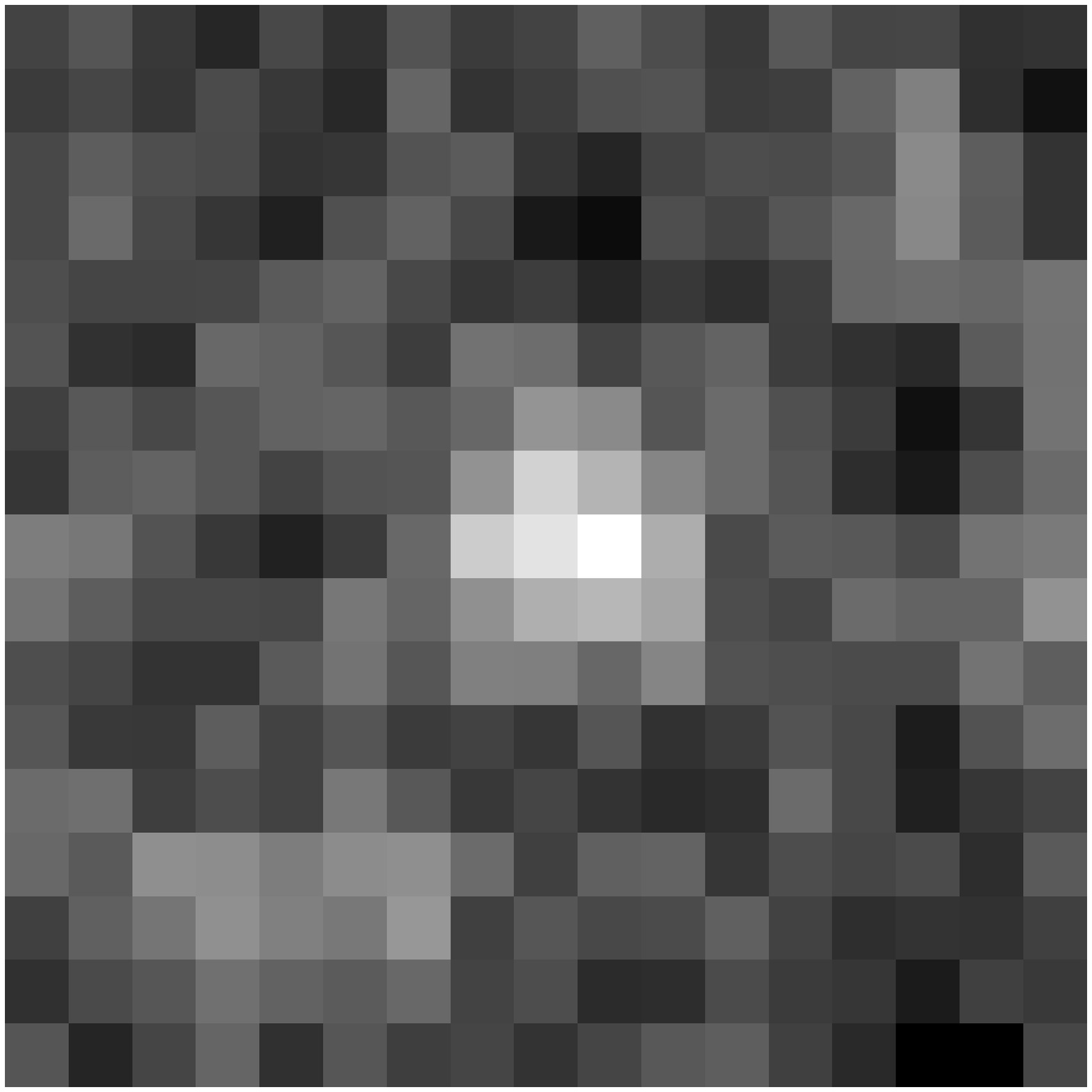}
\includegraphics[scale=0.12]{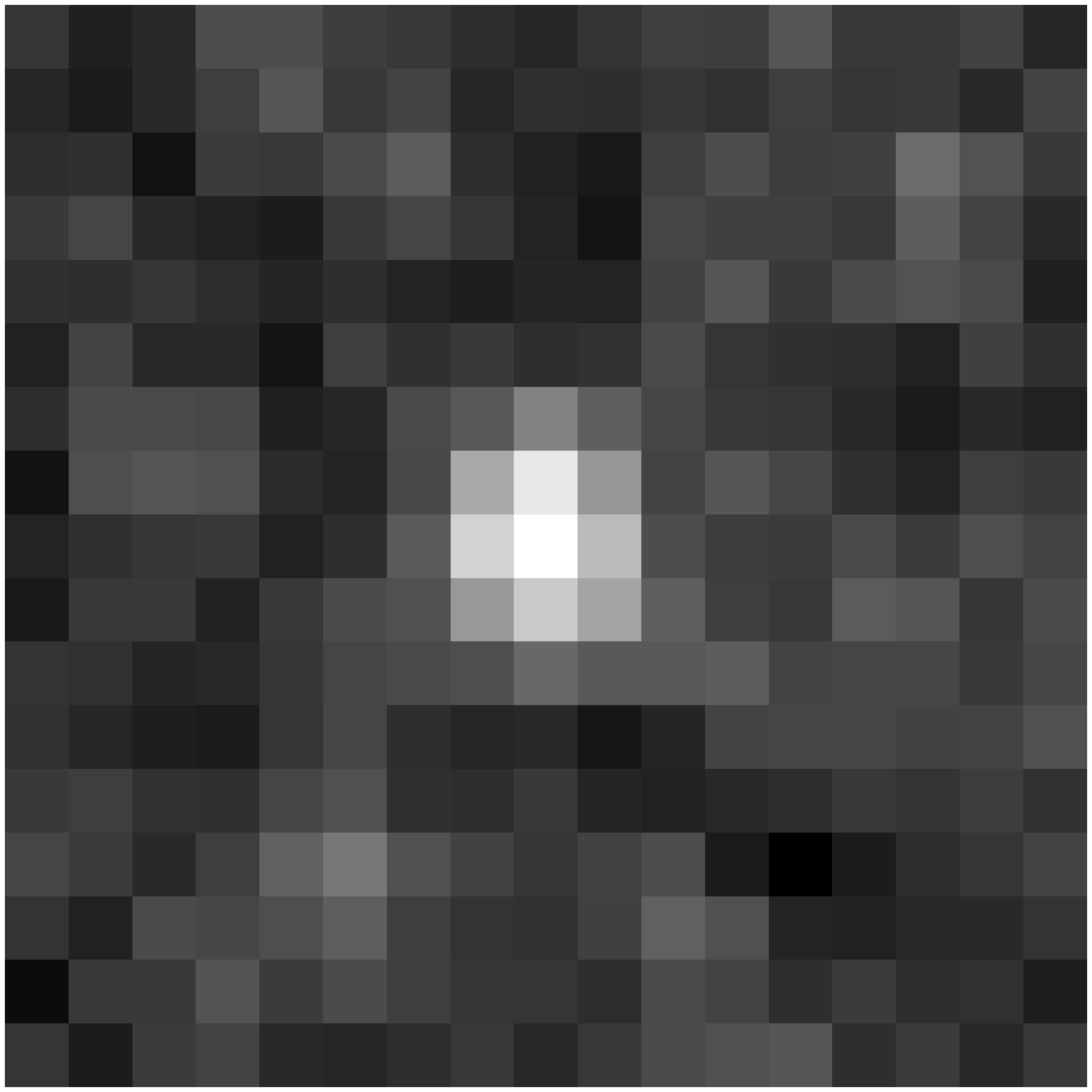}
\end{figure}

\begin{figure}
\centering
\caption{Stacked images from both white dwarfs (left image) and galactic subdwarfs (right), from FIRST cutouts at positions from 2QZ.}
\label{fig:noise}
\includegraphics[scale=0.15]{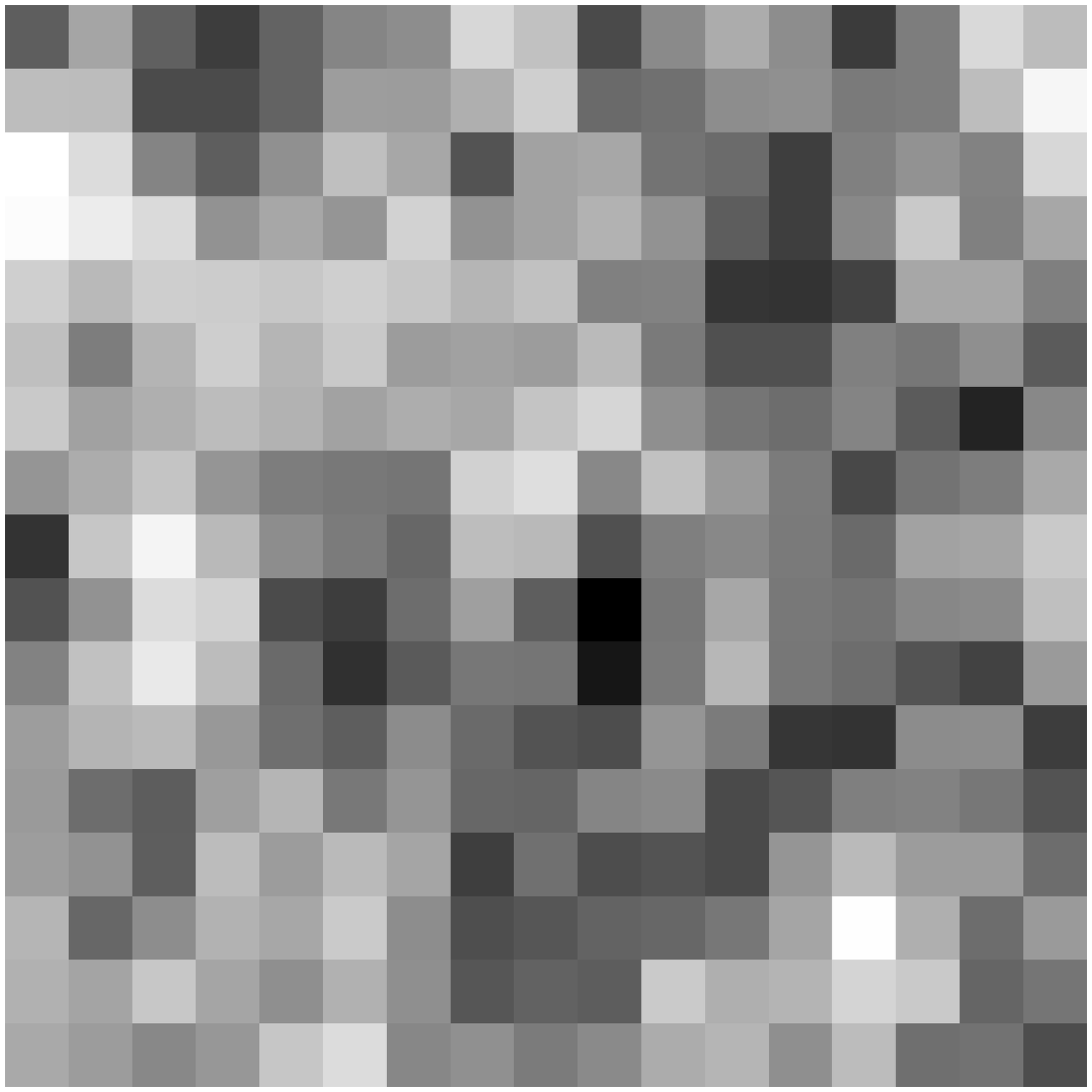}
\includegraphics[scale=0.15]{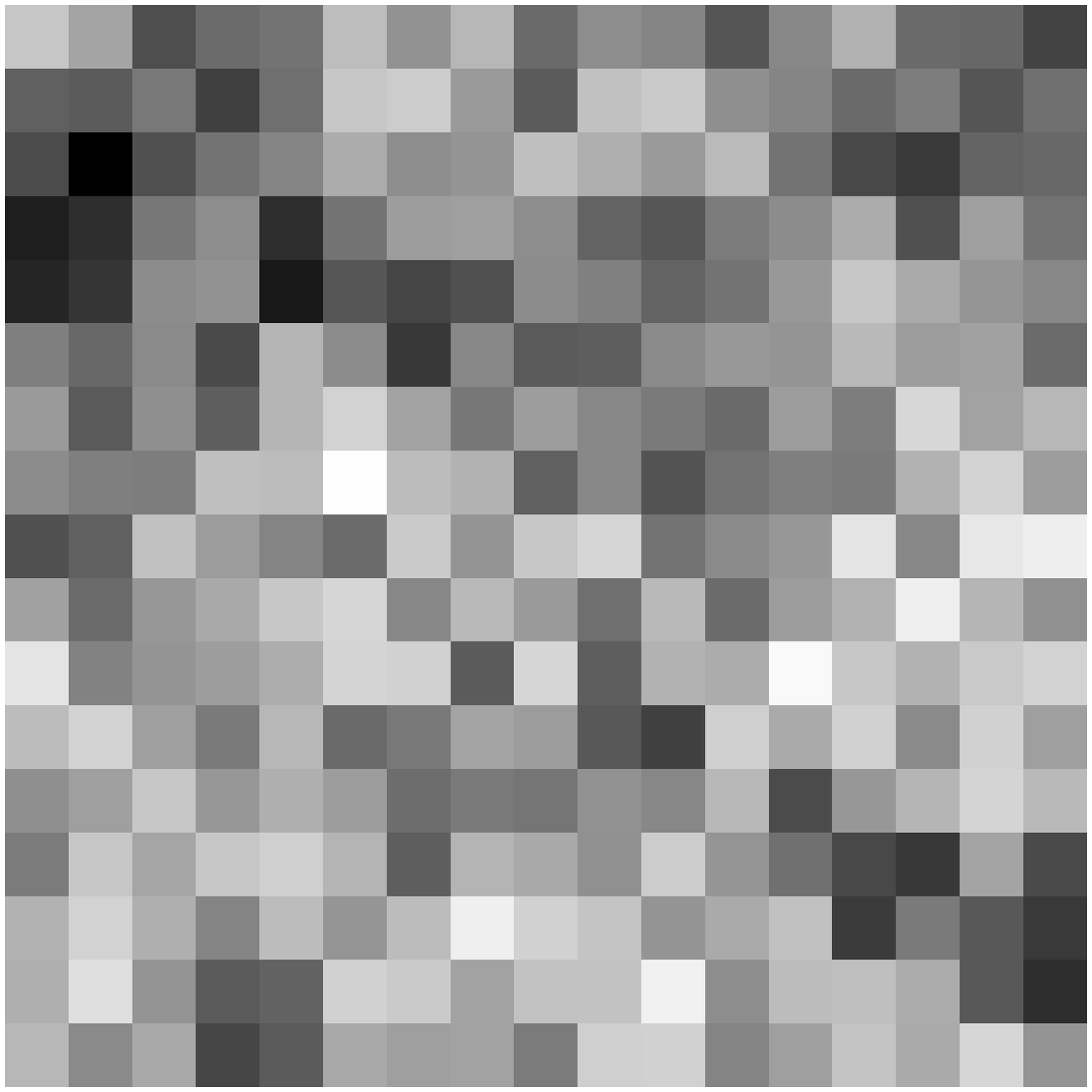}
\end{figure}

The radio images of the QSOs in each redshift bin were stacked using the {\small IRAF} imcombine routine using the median estimator (see Fig.~\ref{fig:bins}).  As a control sample, we also stacked the similar FIRST `cut outs' around the positions of 4830 galactic subdwarfs and 885 white dwarfs in the 2QZ. Galactic subdwarfs are not expected to exhibit radio emission, even at the few $\mu$Jy rms noise levels attained with a stacks of over 4000 0.15mJy-rms images (see Fig~\ref{fig:noise}).

The imfit routine in {\small MIRIAD} was then used to determine the integrated flux of each source detected in the stacked images.  Images were detected at greater than the $5\sigma$ level in each co-added QSO image, but no source detection was made in either the galactic subdwarf or white dwarf control stacks. Images were modelled as point sources, and a gaussian fit was used with a 4-pixel (7.2-arcsec) diameter aperture. This aperture is consistent with the 5-arcsec resolution of the FIRST survey and with the Gaussian 5-arcsec FWHM aperture used to derive the FIRST point-source fluxes used for the radio-loud objects in this analysis. The flux estimate was robust against the choice of the precise centring ($\pm$0.5 pix) of the aperture.  Furthermore, increasing the radius of the aperture produced no significant change in the total derived flux. 

The median QSOs in the redshift bins were found to range in flux from $f(1.4{\rm GHz})_{\rm med}=23\mu$Jy to 42$\mu$Jy.  This point-source flux corresponds to angular scales of between 20\,kpc ($z=0.65$) and 45\,kpc ($z=2.05$) for the median redshift bins used in this analysis.   Given that, the vast majority of quasars is this survey will be compact flat-spectrum sources (see below) this 'point source' flux should represent an unbiased estimate of the total radio flux, independent of redshift.

We note that derived fluxes for these stacked sources are similar to those obtained in the similar analysis made by Glickman et al.\ (2004) on the stacked radio-quiet QSOs in the Sloan QSO survey.

%%%%%%%%%%%%%%%%%

\section {Analysis}

\subsection{Number counts}

\begin{figure}
\includegraphics[scale=0.4]{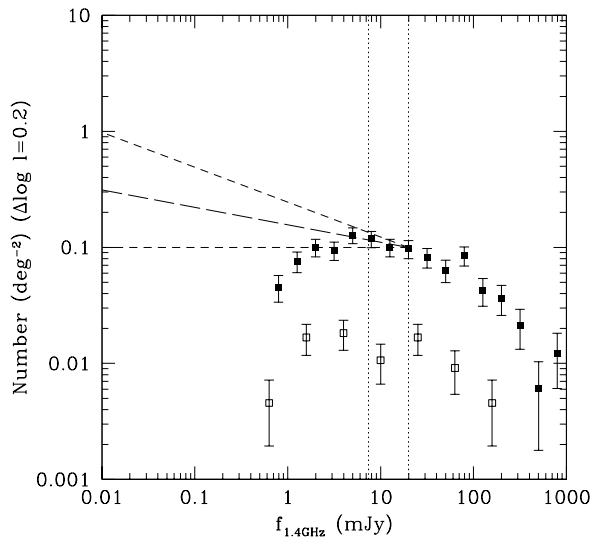}
\caption{Number-flux relation, $n(s)$ for the 2QZ radio-loud (filled squares).  Fits to the $n(s)$ used in the text are indicated by the dashed lines (slopes of 0.0, 0.15 and 0.30 respectively).  The $5\sigma$ completeness limit and $20\,$mJy fitting limit for the FIRST radio survey are shown by the dotted lines. Also shown in the corresponding $n(s)$ relation for the range $1.89<z<2.30$ (open squares).}
\label{plotHist}
\end{figure}

We first calculated the observed radio number-flux relation, $n(s)$, for both the radio-loud and radio-quiet QSOs in the 2QZ to determine whether there is evidence for any bi-modality or discontinuity in the flux distribution which might be indicative of a two-population model. We used the catalogued FIRST 20cm fluxes for the individual radio-loud 2QZ QSOs. The resultant $n(s)$ is shown in Fig~\ref{plotHist}.

We wish to test whether the observed total flux in the stacked images is inconsistent with a smooth extrapolation of the observed radio-loud $n(s)$.   We know that the total flux of the radio-quiet population can not exceed the total flux, $f({\rm 1.4GHz})_{\rm tot}$ obtained by summing the median fluxes detected in each of stacked images over all redshift bins, scaled by the number of QSOs in each stack i.e. $f({\rm 1.4GHz})_{\rm tot} = \sum_{i=1}^{5} N_{\rm QSO}(i) \times f(1.4{\rm GHz})_{\rm med}(i).$  

On the basis that the radio $n(s)$ may be extended below the FIRST completeness limit, we can obtain a prediction for the total QSO radio flux based on a smooth extrapolation of a fit to the radio $n(s)$ above the FIRST completeness limit.  We adopt a completeness limit of 5$\sigma$ (7.5\,mJy) for the FIRST survey, and fit the the $n(s)$ between the 7.5\,mJy and 20\,mJy. Compared to higher fluxes, the slope of $n(s)$ relation for radio-loud QSOs 'flattens-off' at these flux levels. Fitting of the $n(s)$ relation suggest the radio counts may be extrapolated below 7.5\,mJy by a fit ranging from a flat slope, $n(s)$ = constant, to shallow slope $n(s) \propto s^{-0.30}$, with a best-fit $n(s) \propto s^{-0.15}$ (see Fig.~\ref{plotHist}).

Table~\ref{tab:flux} presents the comparison of the integrated flux as a function of depth for these fits between these models and stacked observations. The errors on the integrated flux in both radio-loud domain and radio-quiet domain are approximately 10 per cent. Both model extrapolations converge rapidly at fluxes less than 0.1mJy, with integrated flux predictions for slopes of 0.15 and 0.30 bracketing the observed integrated flux from the 2QZ radio-quiet QSOs. We thus conclude that the flux distribution of radio-quiet QSOs in the 2QZ is not inconsistent with a smooth extrapolation of the 2QZ radio-loud QSO number-flux distribution, on the basis that the slope of the relation flattens at faint flux levels.   

This is a necessary but not sufficient condition for a single population model; more detailed modelling of the source counts based on fitted radio-loud and radio-quiet LFs will be presented elsewhere (Miller et al.\ in preparation). We note that the cosmological luminosity evolution of QSOs in both optical (Croom et al.\ 2004) and radio (Dunlop \& Peacock 2004) tends to cancel out the effects of cosmological dimming over a significant range in redshift ($0.5<z<2$).   This results in luminosity function features such as the break, L*(z), exhibiting the same observed flux over these redshifts.  Thus a simple bimodal luminosity function model with a redshift-dependent break between radio-loud and radio-quiet QSOs that maintains an observed flux close to or below the FIRST radio detection limit over the range in redshifts sampled here could reproduce the current observations.  

We also note that the two-population model favoured by Miller et al.\ (1990) was based on a carefully-selected sample of QSOs at $1.8<z<2.5$. We therefore looked to see if there was any evidence in any of our redshift ranges for the integrated radio-quiet flux below the FIRST completeness limit departing from a smooth extrapolation of the radio-loud $n(s)$.  Due to reduced numbers of objects, the faint slope was much more difficult to constrain in these narrower redshift bins and we retained the values of 0.0 and 0.3 derived for the full $n(s)$ above. The results are reported in Table~\ref{tab:fluxz}, based on an integration from the FIRST 5$\sigma$ limit down to 10$\mu$Jy . As an illustration the $n(s)$ relation for the $1.85<z<2.30$ range is plotted in Fig.~\ref{plotHist}. Although hampered by larger measurement uncertainties, ($\sim 20$ per cent), we found the total integrated observed flux below the FIRST completeness limits ($0.35-0.48\,$mJy deg$^{-2}$) at these redshifts were consistent with our extrapolation of the radio $n(s)$.  In the lowest redshift bin, the total observed flux marginally exceeds the predicted flux with the fitted 0.3 slope, but only at the 1$\sigma$ level.
\begin{table*}
\centering
\caption{Integrated Flux density for observed QSOs and extrapolated $n(s)$}
\label{tab:flux}
\begin{tabular}{@{}rrrrr@{}}
\hline
&\multicolumn{3}{c}{Integrated Flux Density}\\
Flux Limit&$n(s)={\rm const}$&$n(s)\propto s^{-0.13}$&$n(s)\propto s^{-0.2}$&Observed\\
(mJy)&(mJy deg$^{-2}$)&(mJy deg$^{-2}$)&(mJy deg$^{-2})$&(mJy deg$^{-2})$\\
\hline
5.0 &0.50&0.62&0.76&0.64\\
3.2 &0.82&1.03&1.31&0.93\\
2.0 &1.02&1.32&1.71&1.13\\
1.2 &1.14&1.50&1.99&1.23\\
0.8 &1.22&1.63&2.20&1.27\\
. & . & . & . & . \\
. & . & . & . & . \\
%0.5 &1.27&1.72&2.35&1.27\\
%0.3 &1.30&1.78&2.46&1.27\\
%0.2 &1.32&1.82&2.54&1.27\\
%0.13&1.35&1.85&2.60&1.27\\
0.08&1.35&1.87&2.64&1.27\\
0.05&1.35&1.88&2.67&1.52\\
0.03&1.36&1.89&2.70&2.00\\
0.02&1.36&1.90&2.71&2.10\\
0.01&1.36&1.90&2.72&2.10\\

\hline
\end{tabular}
\end{table*}

\begin{table*}
\centering
\caption{Observed and predicted integrated flux densities at f(1.4GHz)$>0.01$ mJy for different redshift bins}
\label{tab:fluxz}
\begin{tabular}{@{}rrrr@{}}
\hline
&\multicolumn{3}{c}{Integrated Flux Density}\\
z range&$n(s)={\rm const}$&$n(s)\propto s^{-0.3}$&Observed\\
&(mJy deg$^{-2}$)&(mJy deg$^{-2}$)&(mJy deg$^{-2})$\\
\hline
$0.12<z<0.85$&0.24&0.40&0.48\\
$0.85<z<1.23$&0.30&0.50&0.40\\
$1.23<z<1.55$&0.32&0.53&0.38\\
$1.55<z<1.89$&0.32&0.53&0.34\\
$1.89<z<2.30$&0.26&0.43&0.38\\
\hline
\end{tabular}
\end{table*}

\subsection{Spectral Index}
\begin{figure}
\includegraphics[scale=0.4]{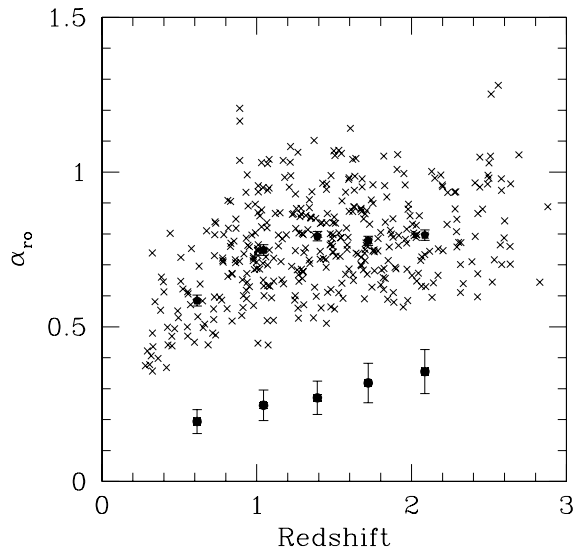}
\caption{Radio to optical spectral index as a function of redshift for both 
radio-loud (crosses) and radio-quiet QSOs (filled circles).}
\label{plotzvsa}
\end{figure}

We also investigated the dependency on redshift of the radio-to-optical luminosity ratio for the radio-loud and radio-quiet QSO populations.  We characterised the radio-to-optical luminosity ratio by $\alpha_{\rm ro}$, the notional spectral index between the radio and optical bands,
$\alpha_{\rm ro} = -(\log {L_{440{\rm nm}}}/{L_{1.4{\rm GHz}}})/(\log {\nu_{440{\rm nm}}}/{\nu_{1.4{\rm GHz}}})$
where $L$ refers to the luminosities in both optical (440nm) and radio (1.4GHz) bands.  Luminosities were calculated using a spectral index of $-0.5$ in both the optical and radio bands.  The assumption that the mean radio and optical spectral indices are independent of the radio `loudness' is implicit in our calculations. Studies of composite optical spectra for radio-loud and radio-quiet quasars confirm that the optical spectral index in the range $1200\AA < \lambda 4400\AA$ is very similar for both populations (Cristiani \& Vio 1990, Zheng et al.\ 1997).  Although there is no information of the radio spectral index of radio-quiet QSOs, at the redshifts under consideration here, evolutionary models of the radio source population suggest that the vast majority of the objects will be flat radio spectrum QSOs largely independent of redshift (see e.g. Jackson \& Wall 1999); steep spectrum sources only dominating at low redshifts $z<0.5$. 

\begin{figure}
\begin{centering}
\includegraphics[scale=0.3]{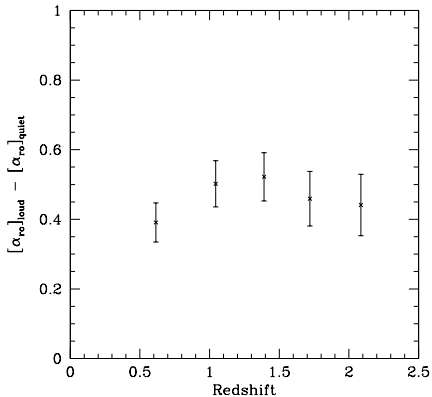}
\caption{The difference between $[\alpha_{ro}]_{RL}$ and  $[\alpha_{ro}]_{RQ}$, as a function of redshift.}
\label{alphaDiff}
\end{centering}
\end{figure}

In Fig~\ref{plotzvsa} the derived $\alpha_{\rm ro}$ for radio-loud and the composite radio-quiet QSOs are shown. Although the absolute redshift trend in $\alpha_{\rm ro}$ is dominated by the choice of spectral index, the relative trend between the radio-loud and radio-quiet QSOs (which is independent of radio or optical spectral index) is similar.  This is confirmed in Figure~\ref{alphaDiff} where we plot the difference in $\alpha_{\rm ro}$ between the radio-loud and radio-quiet QSOs as a function of redshift.  Error bars are determined from the errors on the flux determinations in both optical and radio.  
We can express the relation thus:

$${{L_{RQ}}(z)\over {P_{RQ}(z)}} = A . {{L_{RL}}(z)\over{P_{RL}(z)}}$$

where the subscripts refer to the radio-loud (RL) and radio-quiet (RQ) populations respectively and, for clarity, we denote the optical luminosity ($L_{440\rm nm}$) and radio luminosity ($L_{1.4\rm GHz}$) as $L$ and $P$ respectively.

Over the redshift range of interest ($z<2.3$) the observed radio flux evolution of the radio-loud population and the the optical flux evolution of the radio-quiet populations can be parameterised by luminosity evolution model as follows:

$$L_{RQ}(z) = L^*10^{k_1z+k_2z^2}$$

and 

$$P_{RL}(z) = P^*10^{k'_1z+k'_2z^2}$$

where the optical evolution fit to the radio-quiet population yields $k_1=1.39$ and $k_2=-0.29$ (Croom et al.\ 2004), similar to the radio evolution fit to radio-loud QSOs $k_1=1.18$ and $k_2=-0.28$ (Dunlop \& Peacock 1990).  By combining these relations and expressing the functional forms for the optical and radio evolution as $f_{\rm opt}(z)$ and $f_{\rm rad}(z)$ respectively, we obtain:

$$P_{RQ}(z). L_{RL}(z) = {\rm const} . f_{\rm rad}(z) . f_{\rm opt}(z)$$

We can use the observation by Padovani (1993) that the optical evolution of radio-loud and radio-quiet QSOs are identical ($L_{RL}(z) = L_{RQ}(z)$) to yield:

$$P_{RQ}(z) = {\rm const} . f_{\rm rad}(z)$$

i.e. that the radio evolution of radio quiet QSOs is identical to the radio evolution of radio loud QSOs.  As indicated above, this is predicated on the assumption that the mean optical and radio spectral indices of QSOs are independent of their radio-loudness.  However, if this were not the case, the observation result would itself rely on some fine tuning i.e. the relative difference in spectral indices between radio-loud and radio-quiet QSOs would have to be equal and in the opposite sense to the difference in evolution between the radio-quiet and radio-loud populations. 

This is consistent with the results of Glickman et al.\ (2004) who found that the radio-to-optical luminosity ratio of the stacked radio-quiet QSOs in the SDSS sample exhibited the same dependence on redshift as the radio-loud SDSS QSOs.

\section{Conclusions}
We have demonstrated that the co-addition of blank fields in the FIRST survey provides a robust detection of an ensemble average of radio-quiet QSO flux distribution below the FIRST threshold, and that the flux properties are consistent with a smooth extrapolation of the radio-loud QSO flux distribution detected above this threshold. 

It would be straightforward to apply this technique to other extensive deep optical surveys, (2dFGRS, SDSS galaxies) comprising tens to hundreds of thousands of objects, thus potentially moving into the sub-$\mu$Jy detection limit. With the advent of next generation wide-field radio facilities such as the Square Kilometre Array, the nanoJansky detection limits proposed for surveys could be similarly transformed into sub-nanoJansky limits using these stacking techniques, providing the confusion limit is not reached and suitable object catalogues exist at deep levels in other passbands.

\section*{Acknowledgments}
MW thanks the ATNF for the award of a vacation scholarship during which this work was carried out.

\end{document}